%% file: root.tex
\documentclass[letterpaper, 10 pt, conference]{ieeeconf}  %

\IEEEoverridecommandlockouts                              %

\usepackage{graphics} %
\usepackage{epsfig} %
\usepackage{times} %

\usepackage{amsthm}
\usepackage{amsmath} %
\usepackage{amssymb}  %
\usepackage{mathtools}
\usepackage[short]{optidef}
\usepackage{tensor}
\usepackage{units}
\usepackage{tikz}
\usetikzlibrary{decorations.pathreplacing,calc}
\usepackage{pgfplotstable}
\pgfplotsset{compat=newest}
\usetikzlibrary{calc}
\usepgfplotslibrary{groupplots}
\usepgfplotslibrary{colormaps}
\usepackage{url}

\usepackage{relsize}
\usepackage{subfig}
\usepackage{booktabs}
\usepackage{colortbl}

\definecolor{CCPSblue1}{RGB}{37,64,97}
\definecolor{CCPSblue2}{RGB}{56,96,146}
\definecolor{CCPSblue3}{RGB}{220,230,242}
\definecolor{CCPSgray3}{RGB}{242,242,242}
\definecolor{CCPSgreen1}{RGB}{67,80,58}
\definecolor{mlred}{RGB}{208,2,27}
\definecolor{MPIgreen}{RGB}{0,118,117}
\definecolor{d3_blue}{HTML}{1f77b4}
\definecolor{d3_orange}{HTML}{ff7f0e}
\definecolor{d3_green}{HTML}{2ca02c}
\definecolor{plant}{rgb}{0.75,0.75,0}
\definecolor{nominal}{rgb}{1,0,0}
\definecolor{gp}{rgb}{0,0,1}
\definecolor{bnn}{rgb}{0.75,0,0.75}
\definecolor{constraints}{rgb}{0,0.5,0}

\newcommand{\localBox}[3]
{
\def\Xsize{#1}
\def\Ysize{#2}
\ifnum#3=1
  \draw[fill=MPIgreen!100, opacity=.1] (0,0) rectangle (\Xsize,\Ysize);
  \foreach \x in {0,1,...,\Xsize}{\draw [opacity=1, CCPSgreen1!80] (\x, 0) node[below] {\tiny \x} -- (\x, \Ysize);}
  \foreach \x in {0.5,1.5,...,\Xsize}{\draw [opacity=1, CCPSgreen1!80, dashed] (\x, 0) node[below] {\tiny \x} -- (\x, \Ysize);}
	
  \foreach \y in {0,1,...,\Ysize}{\draw [opacity=1, CCPSgreen1!80] (0, \y) node[left] {\tiny \y} -- (\Xsize, \y);}
  \foreach \y in {0.5,1.5,...,\Ysize}{\draw [opacity=1, CCPSgreen1!80, dashed] (0, \y) node[left] {\tiny \y} -- (\Xsize, \y);}
\else
  \draw[opacity=0] (0,0) rectangle (\Xsize,\Ysize);
  \foreach \x in {0,1,...,\Xsize}{\draw [opacity=0, CCPSgreen1!80] (\x, 0) node[below] {\tiny \x} -- (\x, \Ysize);}
  \foreach \x in {0.5,1.5,...,\Xsize}{\draw [opacity=0, CCPSgreen1!80, dashed] (\x, 0) node[below] {\tiny \x} -- (\x, \Ysize);}
	
  \foreach \y in {0,1,...,\Ysize}{\draw [opacity=0, CCPSgreen1!80] (0, \y) node[left] {\tiny \y} -- (\Xsize, \y);}
  \foreach \y in {0.5,1.5,...,\Ysize}{\draw [opacity=0, CCPSgreen1!80, dashed] (0, \y) node[left] {\tiny \y} -- (\Xsize, \y);}
\fi
}

\theoremstyle{definition}

\DeclareMathAlphabet{\mathantt}{OT1}{antt}{li}{it}

\newcommand{\Tr}{\ensuremath{\mathsf{\mathsmaller T}}}
\newcommand{\norm}[1]{\left\lVert#1\right\rVert}

\title{\LARGE \bf Stochastic Model Predictive Control Utilizing Bayesian Neural Networks}
\author{J. Pohlodek$^{1}$, H. Alsmeier$^{1}$, B. Morabito$^{2}$, C. Schlauch$^{3}$, A. Savchenko$^{1}$, and R. Findeisen$^{1}$%
\thanks{$^{1}$  Control and Cyber-Physical Systems Laboratory, Technical University of Darmstadt, email: { rolf.findeisen@iat.tu-darmstadt.de}}%
\thanks{$^{2}$ Yokogawa Insilico Biotechnology GmbH, Stuttgart}
\thanks{$^{3}$ Humboldt-Universit{\"a}t zu Berlin}
}

\begin{document}

\maketitle
\thispagestyle{empty}
\pagestyle{empty}

\begin{abstract}
Integrating measurements and historical data can enhance control systems through learning-based techniques, but ensuring performance and safety is challenging. Robust model predictive control strategies, like stochastic model predictive control, can address this by accounting for uncertainty. Gaussian processes are often used but have limitations with larger models and data sets. We explore Bayesian neural networks for stochastic learning-assisted control, comparing their performance to Gaussian processes on a wastewater treatment plant model. Results show Bayesian neural networks achieve similar performance, highlighting their potential as an alternative for control designs, particularly when handling extensive data sets.
\end{abstract}

\section{Introduction}
\label{sec:introduction}

Optimal, flexible, safe, and reliable control close to the feasibility boundary is becoming increasingly important to achieve many processes' economic, energy-efficient, and sustainable operation. Optimization-based control, such as model predictive control (MPC)~\cite{rawlings2009,findeisen2002}, is, in principle, well suited to achieve these objectives, as it allows the formulation of elaborate control goals while incorporating state and input constraints. 
However, the MPC performance heavily depends on the quality of the underlying process model~\cite{rawlings2009,forbes2015model}, as it is employed to predict the system's behavior and determine the optimal control action. This brings forward the question of counteracting inevitable model-plant mismatch and measurement uncertainties in real systems. Though nominal MPC exhibits inherent robustness properties~\cite{rawlings2009,findeisen2007sampled} through repeated optimization in the closed loop, uncertainties degrade its performance and can potentially lead to constraint violations or stability loss \cite{rawlings2009}.

Learning-supported MPC approaches aim to reduce the uncertainty in model dynamics by augmenting it with data-driven parts.
However, process uncertainty and stochasticity are often not considered in the MPC formulation if combined with learning approaches; one assumes that the learned part adjusts and compensates for these uncertainties.

In contrast, robust and stochastic MPC approaches \cite{rawlings2009,mesbah2016,mesbah2014stochastic} are explicitly designed for such a scenario, incorporating the information on model uncertainties directly into the MPC formulation. These formulations typically trade off some performance for guaranteeing constraint satisfaction under assumptions on the uncertainty, which is especially important in safety-critical scenarios.

Combining the two strategies --- robust model predictive control strategies and learning --- has been a subject of recent work \cite{bradford2021,hewing2018,hewing2020,maiworm2018stability,morabito2022}. The learned model is often expected to provide some measure of the uncertainty, which is then delivered to the robust or stochastic MPC.   Gaussian processes (GPs) have been widely employed in this role, as they explicitly yield a posterior variance along with the regression mean \cite{Rasmussen2006}. Though GPs are generally easy to tune, their computational complexity grows cubicle with the dataset size, restricting their application to relatively small sets \cite{Rasmussen2006}

An alternative is the use of Bayesian neural networks (BNNs) designed to provide a measure of uncertainty besides the regression \cite{jospin2022}. Belonging to the family of deep learning models, it is computationally efficient, especially as most of the computational effort is spent offline during training. However, compared to other deep learning methods, BNNs are not yet as widely researched --- it is generally challenging to infer the distributions analytically. Instead, one relies on posterior approximations \cite{jospin2022}.

In this work, we aim to determine the suitability of BNNs for (stochastic) predictive control applications, with a particular focus on comparing them to state-of-the-art GPs. BNNs have been used in combination with model predictive control, e.g., \cite{cursi2021bayesian} employs BNNs in combination with a hierarchical MPC approach for the control of a surgical robot to reduce the uncertainty. Here, the kinematics and dynamics of a highly nonlinear robotic system were modeled with the help of BNNs. The uncertainty information provided by the BNN is used in a hierarchical MPC scheme, achieving superior performance. In \cite{bao2022learning}, a learning-based adaptive-scenario-tree model predictive control (MPC) approach is used to achieve probabilistic safety guarantees using BNNs to learn the model uncertainty.

We produced all results in this work using the open-source Python toolbox HILO-MPC\footnote{\small\url{https://www.ccps.tu-darmstadt.de/research_ccps/hilo_mpc/}} \cite{pohlodek2022hilompc}. In a simple-to-use way, the toolbox allows combining (robust) predictive, and optimization-based control and estimation approaches with methods from machine learning, such as Gaussian processes and Bayesian neural networks. \footnote{The case study used as an example, including all code, will be made freely available in HILO-MPC.}

\section{Problem Formulation and Methods}
\label{sec:problem_setup}
We consider dynamic systems which can be described by
\begin{equation}
    x_{k+1}=f(x_k,u_k)+B\left(d\left(x_k,u_k\right)+w_k\right).\label{eq:hybrid_model}
\end{equation}
Here, $x\in\mathbb{R}^{n_x}$ are the dynamical states, $u\in\mathbb{R}^{n_u}$ are the inputs to the system and $B\in\mathbb{R}^{n_x\times n_d}$ is a known matrix. The function $f\colon\mathbb{R}^{n_x}\times\mathbb{R}^{n_u}\rightarrow\mathbb{R}^{n_x}$ describes the known part of the system dynamics, while the function $d\colon\mathbb{R}^{n_x}\times\mathbb{R}^{n_u}\rightarrow\mathbb{R}^{n_x}$ describes an unknown or difficult to model effect. This unknown effect will be learned using specific machine learning algorithms. The variable $w\in\mathbb{R}^{n_d}$ is assumed to be zero-mean normally distributed process noise $w\sim\mathcal{N}\left(0,\Sigma^w\right)$ with the variance matrix $\Sigma^w=\operatorname{diag}\left({\begin{bmatrix}\sigma_1^2 & \dots & \sigma_{n_d}^2\end{bmatrix}}\right)$.
We focus on Gaussian processes and Bayesian neural networks to learn the unknown effect, which we briefly introduce in the following.
\paragraph*{Gaussian Process}
\label{subsec:gp}
We briefly review the main concepts of Gaussian processes. For a more detailed introduction to GPs, we refer to \cite{Rasmussen2006}. A GP is a stochastic supervised machine learning algorithm used in regression and classification tasks. GPs are less prone to overfitting and naturally provide uncertainty measures on predictions. We formulate a regression task as a mapping $\psi\colon\mathbb{R}^{n_\chi}\rightarrow\mathbb{R}:\chi\rightarrow\varphi\left(\chi\right)+\nu$ with the input vector $\chi\in\mathbb{R}^{n_\chi}$, the output $\psi\in\mathbb{R}$ and the zero-mean normally distributed noise $\nu\sim\mathcal{N}\left(0,\sigma^2\right)$ affecting the output. The unknown function $\varphi$ is assumed to be normally distributed $\varphi\left(\chi\right)\sim\mathcal{N}\left(m\left(\chi\right),k\left(\chi,\chi\right)\right)$, with the mean function $m\colon\mathbb{R}^{n_\chi}\rightarrow\mathbb{R}$ and covariance function $k\colon\mathbb{R}^{n_\chi}\times\mathbb{R}^{n_\chi}\rightarrow\mathbb{R}$. These functions are assumed to depend on hyperparameters, that are determined by GP training procedure, resulting in scalar-valued predictions. For simplicity, multiple outputs are handled as separate one-dimensional GPs independently.

\begin{figure}
    \centering
    \begin{tikzpicture}[>=latex]
        \def\helpBox{0}
        \node (HEAD) at (-7.5,4) [anchor=north west,inner sep=0pt] %
        {
        \input{figures/bnn.tikz}
        };
    \end{tikzpicture}
    \caption{a) Bayesian neural network with two features (inputs), one hidden layer, and one label (output). 
    b) Single neuron.
    \\[-5ex]}
    \label{fig:bnn_block_diagram}
\end{figure}
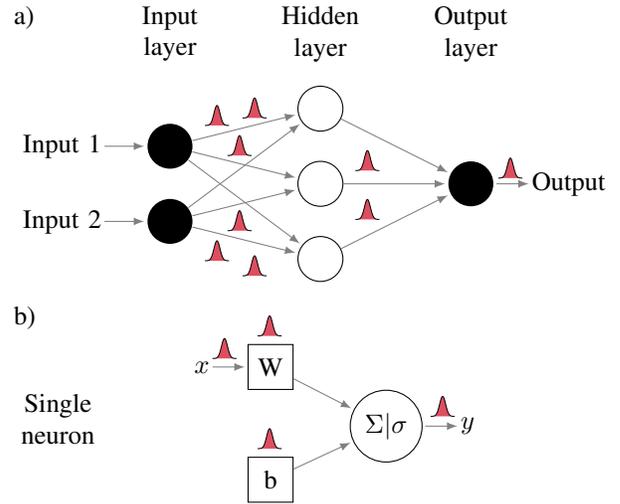

\paragraph*{Bayesian Neural Network}
As a second learning method, we consider Bayesian neural networks \cite{Neal1996}, which can also provide a measure of uncertainty similar to GPs. In contrast to GPs, computational complexity of BNNs does not grow with the number of data points. Unlike traditional feed-forward NNs, BNNs learn distributions over the output instead of single value predictions. To accomplish this task, the weights of conventional neural networks are replaced by distributions over the weight in each layer. Thus, BNNs extend the class of NNs to estimate posterior distributions instead of most likely values, cf. Fig.~\ref{fig:bnn_block_diagram}.

A common way of implementing the weights $\mathcal{W}$ as distributions is to assume that they are independently Gaussian distributed with zero mean and some arbitrary variance $\lambda$ \cite{jospin2022} as follows
\begin{equation}
    p(\mathcal{W})=\prod^{L}_{l=i} \prod^{V_{l}}_{i=0} \prod^{V_{l-1}+1}_{j=1}\mathcal{N}(w_{i,j}|0,\lambda^{-1}).
\end{equation}
Here $L$ denotes here the number of layers in the network and $V_l$ the nodes in each layer $l$. The likelihoods with regard to the network weights are given by
\begin{equation}
    p(y|\mathcal{W},\mathcal{Z}) = \prod^{M}_{i=1}\mathcal{N}(y_i|d(z_i|\mathcal{W}),\gamma^{-1}),
\end{equation}
where $\gamma$ is the precision and $\mathcal{Z}$ is the collection of nodes $z_i$.
Now we can estimate the posterior distribution over the weights given the data via Bayes' rule
\begin{equation}
    p(\mathcal{W}|\mathcal{D}) = \frac{p(y|\mathcal{W},\mathcal{Z})p(\mathcal{W})}{p(y|\mathcal{Z})},\label{eq:bnn_posterior}
\end{equation}
where we can calculate the marginal likelihood via 
\begin{equation}
    p(y|\mathcal{Z}) = \int p(y|\mathcal{W},\mathcal{Z}) \mathcal{W} d\mathcal{W}. \label{eq:marg_like}
\end{equation}
In reality, the marginal likelihood and equation \eqref{eq:marg_like} cannot be calculated, it becomes intractable because of the nonlinearities in the network caused by the activation $d(\cdot,\mathcal{W})$. This results in the need to approximate the posterior distribution. This is a common challenge in the Bayesian inference domain, and different methods exist to approximate the needed posterior. The most popular are Markov chain Monte Carlo (MCMC), Laplace approximation (LA), probabilistic backpropagation (PBP), and variational inference (VI), for a survey see \cite{abdar2021} and \cite{daxberger2021}.

\paragraph*{Nominal Model Predictive Control}
\label{subsec:MPC}
As a comparison and baseline, we consider a nominal model predictive controller, which solves a finite-horizon optimal control problem  at the  sampling times~\cite{rawlings2009,findeisen2002} based on the nominal model\footnote{In principle the nominal MPC could also use the learned "nominal" model part, which would be the mean of the predicted state for case of the GP model. This is avoided here, due to space limitations.}:\\[-4
ex]
\begin{mini!}
    {\{u_k\}}{J(\{x_k\},\{u_k\})}{\label{eq:mpc}}{}
    \addConstraint{x_{k+1}}{=f\left(x_k,u_k\right), \quad x_0=\tilde{x}_j}{}
    \addConstraint{x_k}{\in\mathcal{X}, \quad u_k}{\in\mathcal{U},}{}
\end{mini!}
where $J(\{x_k\},\{u_k\})=\sum_{k=0}^{N}L\left(x_k,u_k\right)+E\left(x_k\right)$ is the cost function, $N$ is the control horizon, $L\colon\mathbb{R}^{n_x}\times\mathbb{R}^{n_u}\rightarrow\mathbb{R}$ is the stage cost, $E\colon\mathbb{R}^{n_x}\rightarrow\mathbb{R}$ is the terminal cost, $\mathcal{X}$ and $\mathcal{U}$ are the feasible input and state sets.
The first input $u_k^\star$ of the obtained optimal input sequence is applied. The optimization problem is repeated at every sampling time updating $x_k$ with the measured state $\Tilde{x}_j$.

\paragraph*{Stochastic Model Predictive Control}
To exploit the stochastic uncertainty information of the model, we use stochastic MPC. Compared to the nominal case, we consider chance constraints, allowing for a violation of the constraints with a certain probability \cite{mesbah2016}. Assuming a cost that contains the expected value of the now stochastic state and control variables minimizing a sequence of optimal policies $\Pi(x) =\{ {\pi_0(x),...,\pi_N(x)}\}$ instead of the optimal open-loop inputs, one obtains the following stochastic optimal control problem
\begin{mini!}
    {\Pi\left(x\right)}{\mathbb{E}\left[J(\{x_k\},\{u_k\})\right]}{}{}
    \addConstraint{}{x_{k+1}=f_\text{prob}\left(x_k,u_k\right),\quad x_0=\tilde{x}_j}{}
    \addConstraint{}{p\left(x_k\in\mathcal{X}\right)\geq p_x,\quad \forall k\in0,\,\dots,\,N}{}
    \addConstraint{}{p\left(u_k\in\mathcal{U}\right)\geq p_u,\quad \forall k\in0,\,\dots,\,N}{}
    \addConstraint{}{u_k=\pi\left(x_k\right).}{}
\end{mini!}
with $f_\text{prob}$ defining the probabilistic model of the system, $p_x$ and $p_u$ are the probabilities with which the chance constraints are allowed to be violated and $k$ describes the time steps of the discrete model. 

Since the stochastic optimal control problem is generally intractable, we reformulate and approximate it to make it tractable and include the learning-based hybrid model. The problem's intractability results from the fact that we need to consider infinitely many state trajectories if we solve for an optimal sequence of policies $\Pi$. 

The reformulations and approximations are based on the approximation presented in \cite{hewing2020} and the theory of reachable sets presented in \cite{hewing2018}. 
We omit the necessary assumptions for the sake of brevity and state the derived outcome: the system state, control input, and uncertainty dynamics, that is to be learned by a GP or a BNN, are jointly Gaussian distributed:
\begin{align}
    \mathcal{N}(\mu_k,\Sigma_k) = \mathcal{N}\left(
     \left[\begin{smallmatrix}
        \mu^x_k \\ \mu^u_k \\ \mu^d_k
    \end{smallmatrix}\right]
    ,
    \left[\begin{smallmatrix}
        \Sigma^x_k & \Sigma^{xu}_k & \Sigma^{xd}_k \\
        \Sigma^{ux}_k & \Sigma^{u}_k & \Sigma^{ud}_k \\
        \Sigma^{dx}_k & \Sigma^{du}_k & \Sigma^{d}_k + \Sigma^{\omega}_k
    \end{smallmatrix}\right]
    \right).
    \label{eq:jointdis}
\end{align}
Furthermore, we assume that the chance constraints are tightening around the mean of our input and state variables $\mu^x$ and $\mu^u$, which are defined by the assumed Gaussian distributions. With these assumptions, we can formulate of tractable stochastic optimal control problem of the following structure (as we only want to show the basic structure and due to limited space, we omit introducing all symbols)
\begin{mini!}
    {\mu^u_k}{\mathbb{E}\left[J(\{x_k\},\{u_k\})\right]}{}{}
    \addConstraint{}{\mu^x_{k+1}=\hat{f}\left(x_k,u_k,k\right)+B_d\mu^d_k}{}
    \addConstraint{}{\Sigma^x_{k+1}\!=\!\left[\nabla\hat{f}\left(\mu^x_k,\mu^u_k\right)B_d\right]\!\Sigma\!\left[\nabla\hat{f}\left(\mu^x_k,\mu^u_k\right)B_d\right]^\Tr\!\!}{}
    \addConstraint{}{\mu^x_0=\tilde{x}_j, \mu^d_k=\mu^d\left(\mu^x_k,\mu^u_k\right), \Sigma^x_0=0}{}
    \addConstraint{}{\Sigma_k=\text{according to eq. \eqref{eq:jointdis}}}{}
    \addConstraint{}{\mu^x_k\!\in\!\mathcal{Z}\!\left(\Sigma^x_{k+1}\right)\!, \mu^u_k\!\in\!\mathcal{V}\!\left(\Sigma^x_{k+1}\right),\;\forall k\!\in\!0,\,\dots,\,N}{}
\end{mini!}
This reformulation and approximation allow fast solutions. It is implemented in HILO-MPC \cite{pohlodek2022hilompc} allowing to use GP and BNN models. A more detailed discussion of the case of GPs can be found in \cite{morabito2022}.

\section{Case Study: Wastewater Treatment Plant}
\label{sec:case_study}

We want to compare the suitability of hybrid GP and BNN models for model-based control using stochastic MPC. To do so, a wastewater treatment plant is considered, as introduced in \cite{hu2002}, which can be described by \\[-2ex]
\begin{align*}
    \Dot{X}(t) &=\mu(X,S)X(t)- \frac{F(t)}{V}X(t)-K_dX(t)+w_X(t),\\
    \Dot{S}(t) &=\frac{F(t)}{V}(S_f(t)-S(t))-\frac{\mu(X,S)}{Y}X(t)+w_S(t).\label{eq:substrate_ode}
\end{align*}
Here, the states $(X,S)$ denote the total biomass ($\text{mg}/L$) and the substrate concentrations ($\text{mg}/L$). The mapping $\mu(X,S)$ determines biomass growth rate, while $K_d$ denotes its death rate. The in- and outflow rate is defined by $F(t)$ ($L/d$) over the reactor volume $V$ ($L$), and the substrate is fed with the concentration of $S_f(t)$ ($\text{mg}/L$). The dimensionless parameter $Y$ denotes the substrate yield coefficient. Furthermore, normally distributed process noise is acting on each state, i.e., $w_X\sim\mathcal{N}\left(0,1\right)$ and $w_S\sim\mathcal{N}\left(0,1\right)$.
The reactor is controlled by adjusting the flow rate directly, i.e., $u(t)=F(t)$.
The real specific growth rate $\mu$ is defined by the Contois equation
  $  \mu_\text{con}\left(X,S\right)=\dfrac{\mu_\text{max,con}S}{BX+S},$
where $\mu_\text{max,con}$ is the maximum growth rate and $B$ is the kinetic saturation coefficient.

For the control model, the growth rate is assumed to be a Monod equation leading to a model-plant mismatch. The Monod equation is given by 
    $\mu_\text{mon}=\dfrac{\mu_\text{max,mon}S}{K_S+S},$
with the half-velocity constant $K_S$. Furthermore, we assume a mismatch in the process parameters $K_S$ and $\mu_\text{max,mon}$ of $+10\%$ and $-20\%$, respectively, from the nominal parameters. All parameters used in the true plant and the nominal controller model are given in Table \ref{tab:parameters}.
\begin{table}
    \centering
    \caption{Parameters used in the simulations \cite{hu2002}.}
    \begin{tabular}{|c|c|c|c|c|c|}
        \hline
        \rowcolor{CCPSblue1}
        \color{white} parameters & \color{white} values & \color{white} parameters & \color{white} values & \color{white} parameters & \color{white} values \\
        \hline
        \rowcolor{CCPSblue3}
        $K_d$ & $0.0131$ & $\mu_\text{max,con}$ & $0.9297$ & $\mu_\text{max,mon}$ & 0.6275 \\
        $Y$ & $0.2116$ & $B$ & $0.4818$ & $K_S$ & $443.1$ \\
        \rowcolor{CCPSblue3}
        $V$ & $5$ & $-$ & $-$ & $-$ & $-$ \\
        \hline
    \end{tabular}
    \label{tab:parameters}\\[-2ex]
\end{table}
The influent substrate concentration $S_f\left(t\right)$ is modeled as a time-varying parameter, as typically is the case in real world wastewater plants. For $S_f$, we assume it underlies a smooth and random but bounded disturbance
  $  S_f\left(t\right)=5500+100\left(\sin\left(w_tt\right)+\sin\left(w_\pi\pi t\right)+\sin\left(w_\mathrm{e}\mathrm{e}t\right)\right),\label{eq:influent_substrate}$
with $\mathrm{e}$ being Euler's number and 
    $w_t\sim\mathcal{N}\left(0.3,0.01\right)+\mathcal{N}\left(-0.3,0.01\right),$ $
    w_\pi\sim\mathcal{N}\left(0.01,0.01\right)+\mathcal{N}\left(-0.01,0.01\right), $ $w_\mathrm{e}\sim\mathcal{N}\left(0.08,0.01\right)+\mathcal{N}\left(-0.08,0.01\right).$

All simulations were run on a MacBook Pro with an Apple M2 chip and $24\,\text{GB}$ memory using the operating system macOS Ventura (version 13.2.1). CasADi \cite{Andersson2019}, which is used internally by HILO-MPC, was installed with its most recent version (version 3.5.5). Additionally, the linear solver HSL\_MA97 from the HSL package \cite{hsl} was used.%

\paragraph*{Data Generation \& Training}
\label{subsec:data_generation_and_training}
The true plant model and the nominal control model were used to generate the data for the training of both, the GP and the BNN. To do so, we created a nominal MPC design according to \eqref{eq:mpc}. The stage cost $L$ and the terminal cost $E$ are defined as
   $ L\left(x_k,u_k\right)=\norm{x_k-x_\text{ref}}_{Q_s}^2+\norm{u_k-u_\text{ref}}_{R_s}^2 
    +\norm{u_k-u_{k-1}}_{R_c}^2,
    E\left(x_k\right)=\norm{x_N-x_\text{ref}}_{Q_t}^2,$
with the reference values
    $x_\text{ref}=\begin{bmatrix}
        X_\text{ref} & S_\text{ref}
    \end{bmatrix}=\begin{bmatrix}
        1046.28 & 101.615
    \end{bmatrix}, 
    u_\text{ref}=F_\text{ref}=0.714286,$
and the weights
 $   Q_s=Q_t=\left[\begin{smallmatrix}
        10 & 0 \\ 0 & 10
    \end{smallmatrix}\right],\quad R_s=1,\quad R_c=5\cdot10^3.$
 The reference of the flow rate $F_\text{ref}$ translates to a hydraulic retention time of $\tau=7\,\mathrm{d}$, i.e., the average time a volume of wastewater will remain in a particular part of the plant, and lies within the range of hydraulic retention times referenced in \cite{hu2002}. The reference values of the biomass concentration $X_\text{ref}$ and the substrate concentration $S_\text{ref}$ are the steady states of the open loop simulation using the chosen reference flow rate $F_\text{ref}$. The initial conditions of the states were $X_0=0.2$ and $S_0=0$, and both states were constrained to be at least zero over the whole process time. Furthermore, the input was constrained to lie within the range $0\leq F(t)\leq2$. The sampling interval was $\Delta t=0.125\,\mathrm{d}$ and the length of the control horizon was $N=80$, which translates to a time of $10\,\mathrm{d}$. The influent substrate concentration was calculated using \eqref{eq:influent_substrate} and was kept constant over the control horizon.

We ran six closed loop simulation using this nominal MPC for a simulation time of $T_\text{sim}=70\,\mathrm{d}$ for each simulation, which ensured that the steady state was reached in each simulation. Additionally, we sampled the reference states for each simulation from a uniform distribution around the actual reference states to gain a sufficient distribution of data in the state space region of interest
   $ \Tilde{X}_\text{ref}\sim\mathcal{U}\left(0.9X_\text{ref},1.1X_\text{ref}\right), 
    \Tilde{S}_\text{ref}\sim\mathcal{U}\left(0.9S_\text{ref},1.1S_\text{ref}\right).$
Both states are assumed to be measured at all times and serve as the features of the machine learning models. The discrepancy between the true plant model and the nominal control model, the $d$ part of \eqref{eq:hybrid_model}), is used as the label for the machine learning models.
Overall, $2800$ data points were assembled and split between training and test sets with the ratio $0.8:0.2$.
The data was scaled to zero-mean and standard deviation to improve training performance.

\paragraph*{Gaussian Process}

Since the training of a GP is computationally expensive, a sparser subset of training data was generated. This was done by iterating through all observations in the training data and disregarding all the observations that have an Euclidean distance to the current observation below a certain threshold. This way we were able to obtain more observations from regions that have been observed very little and disregard observations from well-observed regions in the state space. Since the data was normalized, the threshold was set to $0.2$. The new training data set obtained using this threshold amounts to $92$ data points.

\begin{figure}
    \centering
    \input{figures/contour_x.tikz}\vspace{-2ex}
    \caption{Heat map of the predictive standard deviation for the biomass $X$. The black dots are the training data points. }\vspace{-4ex}
    \label{fig:heat_map_x}
\end{figure}
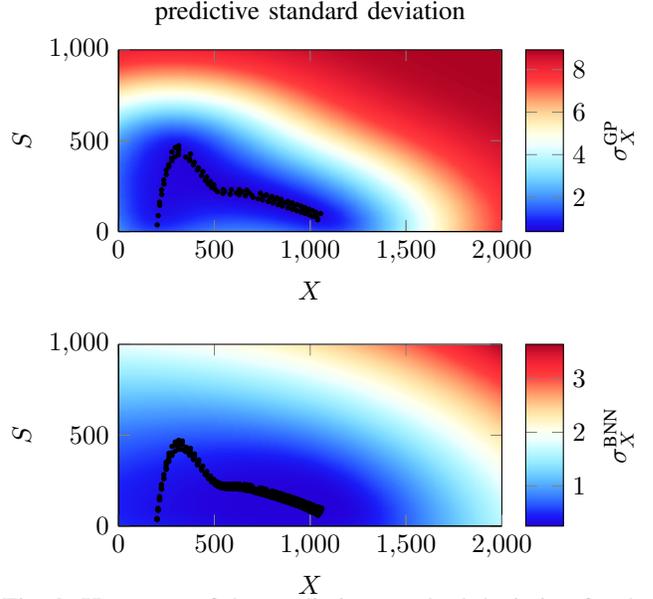

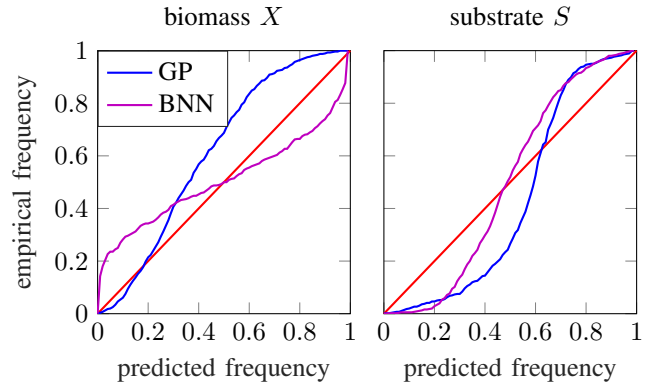
\begin{figure}
    \centering
    \input{figures/miscal.tikz}\vspace{-3ex}
    \caption{Calibration curves, the ideal calibration is shown red.
    } \vspace{-3ex}
    \label{fig:miscalibration_area}
\end{figure}

As the model of the wastewater treatment plant has two states, we trained two GPs. The noise variance of each GP was fixed to $\sigma_X^2=0.1$ and $\sigma_S^2=0.01$, respectively. We chose a squared exponential kernel with automatic relevance detection as the covariance function for both GPs
 $   k\left(\chi_i,\chi_j\right)=\sigma_f^2\exp\left(-\dfrac{\left(\chi_i-\chi_j\right)^2}{2\ell^2}\right),$
where $\sigma_f$ is the signal variance and $\ell$ are the length scales. The mean function was assumed to be zero for both GPs, as is typically done \cite{Kocijan2003}. The hyperparameters of the trained GPs can be found in Table \ref{tab:gp_parameters}.
\begin{table}
    \centering
    \caption{Hyperparameters of the trained GPs.}
    \begin{tabular}{|c|c|c|}
        \hline
        \rowcolor{CCPSblue1}
         & \color{white} length scales $\ell$ & \color{white} signal variance $\sigma_f^2$ \\
        \hline
        \rowcolor{CCPSblue3}
        $\text{GP}_X$ & $\begin{bmatrix} 2.35425 & 2.91528\end{bmatrix}$ & $5.86936$ \\
        $\text{GP}_S$ & $\begin{bmatrix}2.82233 & 2.44933\end{bmatrix}$ & $10.493$ \\
        \hline
    \end{tabular}
    \label{tab:gp_parameters}
\end{table}
The length scales of both GPs are relatively close to each other, indicating that every input to the GPs is equally important. Fig.~\ref{fig:heat_map_x} shows the resulting predictive standard deviation for $\text{GP}_X$ in the top row.
Its model uncertainty is small in the region around the training data points. In regions where there are no observations available, it predicts a very high uncertainty. Fig.~\ref{fig:miscalibration_area} shows the calibration curves of both GPs \cite{kuleshov2018}.
The calibration curve for $\text{GP}_X$ is very close to the ideal calibration line, indicating reliable predictions of the model-plant mismatch $d_X$ for the biomass concentration. This can also be shown by calculating the miscalibration area, i.e., the area between the curve and the ideal calibration line. The closer the calibration curve is to the perfect calibration line, the lower the miscalibration area, resulting in higher prediction reliability. The resulting miscalibration area is listed in Table~\ref{tab:ml_performance}.
\begin{table}
    \centering
    \caption{Root-mean-square error (RMSE), negative log likelihood (NLL) and micalibration area of the trained GPs and BNNs. Bold indicates the lowest value for each metric.}
    \begin{tabular}{|l|l|l|l|}
        \hline
        \rowcolor{CCPSblue1}
         & \color{white} RMSE & \color{white} NLL & \color{white} Miscalibration area \\
        \hline
        \rowcolor{CCPSblue3}
        $\text{GP}_X$ & $0.173$ & $\mathbf{-0.277}$ & $0.117$ \\
        $\text{BNN}_X$ & $\mathbf{0.165}$ & $0.478$ & $\mathbf{0.107}$ \\
        \hline
        \rowcolor{CCPSblue3}
        $\text{GP}_S$ & $\mathbf{0.205}$ & $0.148$ & $0.135$ \\
        $\text{BNN}_S$ & $0.264$ & $\mathbf{0.043}$ & $\mathbf{0.101}$ \\
        \hline
    \end{tabular}
    \label{tab:ml_performance}\vspace{-4
    ex}
\end{table}
The calibration curve for the GP trained on the model-plant mismatch in the substrate concentration ($\text{GP}_S$) is a further away from the ideal calibration line, leading to a higher miscalibration area (see Table~\ref{tab:ml_performance}). This indicates a lower prediction reliability of $\text{GP}_S$ compared to $\text{GP}_X$. Two additional metrics, the root-mean-square error (RMSE) as well as the negative log-likelihood (NLL), are listed in Table~\ref{tab:ml_performance}. The RMSE reflects the correctness of the predictions, and the NLL is another indicator of the predictions' reliability. Both metrics respectively show that $\text{GP}_X$ has higher accuracy and reliability in the forecasts compared to $\text{GP}_S$, as both values are lower for $\text{GP}_X$.

\subsubsection{Bayesian Neural Network}

In this work, we used probabilistic backpropagation \cite{hernandez-lobatoc2015} to approximate the posterior distribution \eqref{eq:bnn_posterior} of the BNN, which is a closed-form approximation based on the assumed density filtering approach
\cite{minka2001}. Its full derivation is beyond the scope of this work, so we refer the reader to \cite{hernandez-lobatoc2015} for details. In contrast to our work, \cite{cursi2021bayesian} and \cite{bao2022learning} employ variational inference to approximate the posterior distribution.

In probabilistic backpropagation, the prior precision $\lambda$ and the noise precision $\gamma$ are assumed to be Gamma distributed hyperpriors
 $   p\left(i|\alpha_i,\beta_i\right)\sim\Gamma\left(\alpha_i,\beta_i\right)$
where $\alpha_i$ are the shape parameters and $\beta_i$ are the inverse scale parameters. Here, we specified the weight hyperpriors with shape $\alpha_\lambda=6$ and inverse scale $\beta_\lambda=6$ for both outputs. The noise hyperpriors were $\alpha_{\gamma,X}=\beta_{\gamma,X}=40$ and $\alpha_{\gamma,S}=\beta_{\gamma,S}=6$, respectively. As with the GPs, the BNNs also need to be trained for each output individually. Both BNNs had one hidden layer with $50$ nodes and the ReLU activation function \cite{hernandez-lobatoc2015}. The BNNs were trained for $10$ epochs each.

Fig.~\ref{fig:heat_map_x} shows the resulting predictive distribution for $\text{BNN}_X$ in the bottom row. Similar to $\text{GP}_X$, it predicts a very low uncertainty in the regions where the training data are located. Unlike $\text{GP}_X$, $\text{BNN}_X$ centers its posterior on the best observed region, shows higher uncertainty in regions with fewer observations, but lower uncertainties in regions without observations. The calibration curves of both BNNs are also very close to the ideal calibration line, indicating high reliability in the predictions. Overall, BNNs and GPs show competitive performance, which is also reflected when comparing the metrics (RMSE, NLL, miscalibration area --- see Table~\ref{tab:ml_performance}).

\begin{figure}
    \centering
    \includegraphics{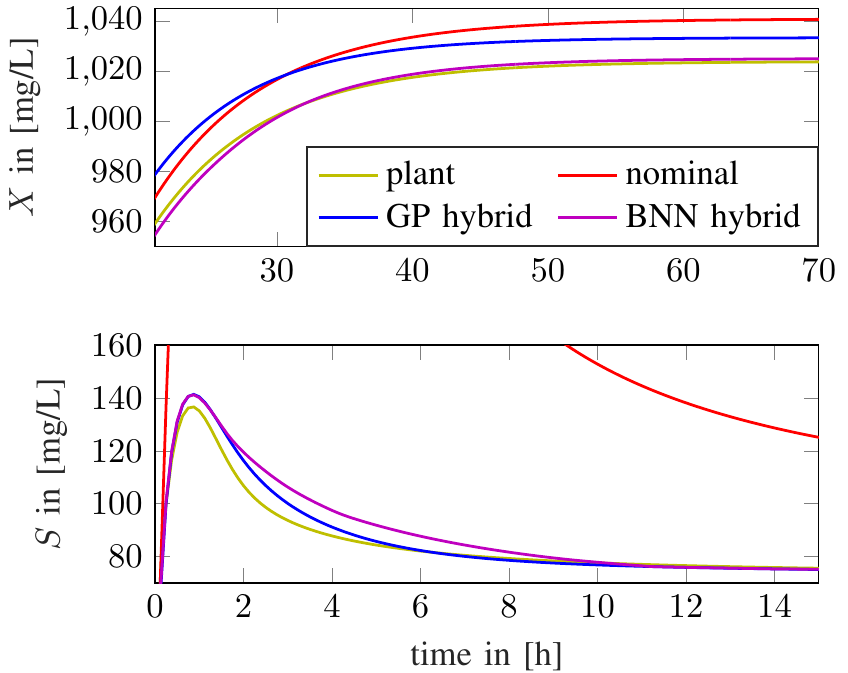}\\[-2ex]
    \caption{Open loop step response of the true plant model.}\vspace{-3ex}
    \label{fig:open_loop}
\end{figure}

\paragraph*{Open Loop Simulation of Hybrid Models}
Given the trained GPs and BNNs, we generated the respective hybrid models. To illustrate their recovery of the model-plant mismatch, we compared open loop step response versus the true plant model and the nominal control model without the process noise. The result for a the step response is shown in Fig.~\ref{fig:open_loop}.
Both hybrid models significantly improve the nominal control model, especially substrate concentration. The performance improvement compared to the nominal control model is substantial, while the mismatch in the steady states of the biomass concentration of a true model is negligible if process noise were considered.

\paragraph*{Closed Loop Control}
Finally, learning-supported stochastic MPC for both learning approaches is evaluated. Box constraints of the nominal MPC augmented with a time-varying constraint that ensures survival of the biomass despite disturbed substrate concentration $S_f$. This constraint is realistic for practical applications that assume model-plant mismatches, since drastic drops in $S_f$ could wash out the biomass and drastic increases of $S_f$ could lead to unmodeled behavior, e.g., cell aggregation. Hence, we demand the biomass concentration to be within once the steady state is reached
 $ X(t)\in [(X_\text{ref}-20) b_l, (X_\text{ref}+20) b_u]$. %
For $t=30\,\mathrm{d}$ after the steady state is reached, the parameters $b_l$ and $b_u$ are set to
   $b_l\left(t\right)=\tanh\left(0.1t+0.01\right)$, 
    $b_u\left(t\right)=\tanh\left(0.1t+1\right).$
For the stochastic MPC we reformulated the constraint %
into a chance constraint with an acceptance probability of $99\%$.
In order to ease the computational burden, the control horizon was shortened to $N=32$ time steps.%

\begin{figure}[hbt]
    \centering
    \input{figures/smpc_hybrid.tikz}\vspace{-1ex}
    \caption{Closed loop simulations.}\vspace{-3ex}
    \label{fig:smpc_hybrid}
\end{figure}
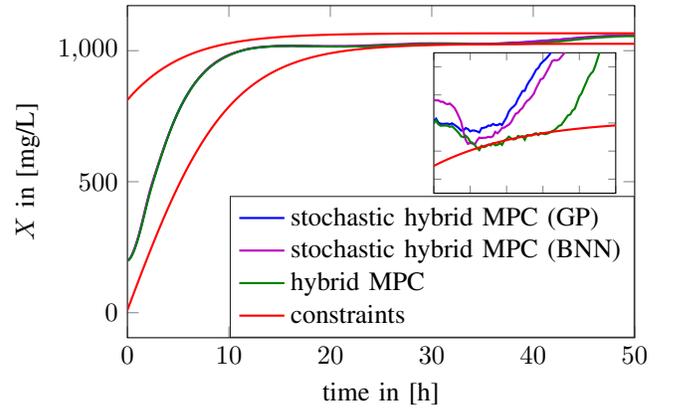
Fig.~\ref{fig:smpc_hybrid} shows the closed loop behavior of the control designs. The difference between the approaches is mainly present in the vicinity of the constraints. Disturbances in $s_f$ lead to learning-supported MPC violating lower constraint, showing its inability to guarantee constraint satisfaction. In contrast, the learning-supported stochastic MPC designs satisfy the constraints.

\begin{figure}[htb]
    \centering
    \input{figures/comp_time.tikz}\vspace{-3ex}
    \caption{Comparison of the total computation time.
    }\vspace{-3ex}
    \label{fig:comp_time}
\end{figure}
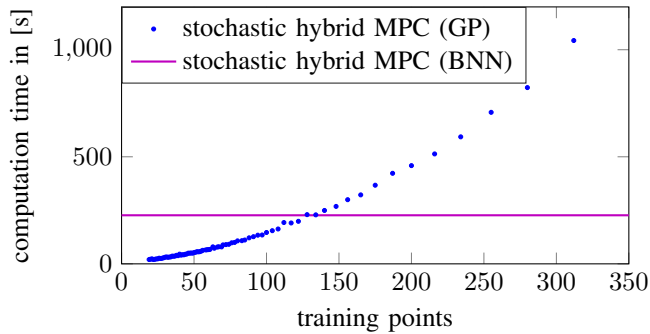
Fig.~\ref{fig:comp_time} compares the computation time for both stochastic MPC controllers.
Multiple GPs were trained and run in the closed loop control setup, varying the number of training points. One can see the increase in total computation time of the corresponding stochastic MPC, and around 130 training data points it surpasses the computation time of the stochastic MPC using the BNN.

\section{Conclusion and Outlook}
\label{sec:conclusion_and_outlook}
State-of-the-art learning-supported stochastic MPC requires uncertainty prediction, often using Gaussian processes. However, their computational complexity increases with data set size. We explored Bayesian neural networks (BNNs) for hybrid models in learning-supported stochastic MPC, comparing their performance to Gaussian processes. BNNs achieved similar performance, minimized model-plant mismatch in open-loop simulations, and proved effective in closed-loop simulations. BNNs offer a valuable alternative, efficiently handling large data sets.
All results used the open-source Python toolbox HILO-MPC \cite{pohlodek2022hilompc}.

Future research will investigate BNNs for more complex models and provide strict stability and performance guarantees for specific model classes.

\addtolength{\textheight}{-8cm}   %

\section*{ACKNOWLEDGMENT}
The authors acknowledge funding of the DIGIPOL project (Magdeburg, Saxony-Anhalt) funded in the EU-ERDF scheme and the KI-Embedded project funded by the BMWK.
\bibliographystyle{IEEEtran}
\bibliography{IEEEabrv, BNN_SMPC}
\end{document}

%% file: figures/bnn.tikz
\begin{tikzpicture}[>=latex]

    \ifnum\helpBox=1, \tikzset{helpBoxOption/.style={draw=CCPSgreen1, line width=1pt}}, \else, \tikzset{helpBoxOption/.style={draw=none}}, \fi
    
    \pgfmathdeclarefunction{gauss}{2}{%
      \pgfmathparse{1/(#2*sqrt(2*pi))*exp(-((x-#1)^2)/(2*#2^2))}%
    }
    
    \node (Box) at (0,0) [anchor=south west, inner sep=0pt, helpBoxOption]
    {
    \begin{tikzpicture}
        \localBox{8}{7}{\helpBox}
        
        \node (imag) at (0,7) [anchor=north west]{
        \def\layersep{2cm}
        \begin{tikzpicture}[shorten >=1pt,->,draw=black!50, node distance=\layersep]
            \tikzstyle{every pin edge}=[<-,shorten <=1pt]
            \tikzstyle{neuron}=[circle,fill=black!25,minimum size=17pt,inner sep=0pt]
            \tikzstyle{input neuron}=[neuron, fill=black];
            \tikzstyle{output neuron}=[neuron, fill=black];
            \tikzstyle{hidden neuron}=[neuron, fill=white, draw=black];
            \tikzstyle{annot} = [text width=4em, text centered]
            \foreach \name / \y in {1,...,2}
                \node[input neuron, pin={[pin distance=0.6cm]left:{Input \y}}] (I-\name) at (0,-\y) {};
            \foreach \name / \y in {1,...,3}
                \path[yshift=0.5cm]
                    node[hidden neuron] (H-\name) at (\layersep,-\y cm) {};
            \node[output neuron,pin={[pin edge={->},pin distance=0.5cm]right:{Output}}, right of=H-2] (O) {};
            \foreach \source in {1,...,2}
                \foreach \dest in {1,...,3}
                    \path (I-\source) edge (H-\dest);
            \foreach \source in {1,...,3}
                \path (H-\source) edge (O);
            \node[annot,above of=H-1,node distance=1cm] (hl) {Hidden layer};
            \node[annot,left of=hl] {Input layer};
            \node[annot,right of=hl] {Output layer};
        \end{tikzpicture}
        };
        \node (gauss_out1) at (6.4,5) [anchor=north west]{
        \resizebox{0.4cm}{!}{
            \begin{tikzpicture}
                \begin{axis}[
                domain=0:8,
                samples=100,
                axis line style={draw=none},
                tick style={draw=none},
                xticklabels=\empty,
                yticklabels=\empty
                ]
                    \addplot[very thick,black,fill=mlred!70] {gauss(4,1)};
                \end{axis}
            \end{tikzpicture}
        }
        };
        \node (gauss_in_out1) at (4.5,5.1) [anchor=north west]{
        \resizebox{0.4cm}{!}{
            \begin{tikzpicture}
                \begin{axis}[
                domain=0:8,
                samples=100,
                axis line style={draw=none},
                tick style={draw=none},
                xticklabels=\empty,
                yticklabels=\empty
                ]
                    \addplot[very thick,black,fill=mlred!70] {gauss(4,1)};
                \end{axis}
            \end{tikzpicture}
        }
        };
        \node (gauss_in_out2) at (4.5,4.45) [anchor=north west]{
        \resizebox{0.4cm}{!}{
            \begin{tikzpicture}
                \begin{axis}[
                domain=0:8,
                samples=100,
                axis line style={draw=none},
                tick style={draw=none},
                xticklabels=\empty,
                yticklabels=\empty
                ]
                    \addplot[very thick,black,fill=mlred!70] {gauss(4,1)};
                \end{axis}
            \end{tikzpicture}
        }
        };
        \node (gauss_hidden1) at (2.5,5.7) [anchor=north west]{
        \resizebox{0.4cm}{!}{
            \begin{tikzpicture}
                \begin{axis}[
                domain=0:8,
                samples=100,
                axis line style={draw=none},
                tick style={draw=none},
                xticklabels=\empty,
                yticklabels=\empty
                ]
                    \addplot[very thick,black,fill=mlred!70] {gauss(4,1)};
                \end{axis}
            \end{tikzpicture}
        }
        };
        \node (gauss_hidden2) at (3,5.8) [anchor=north west]{
        \resizebox{0.4cm}{!}{
            \begin{tikzpicture}
                \begin{axis}[
                domain=0:8,
                samples=100,
                axis line style={draw=none},
                tick style={draw=none},
                xticklabels=\empty,
                yticklabels=\empty
                ]
                    \addplot[very thick,black,fill=mlred!70] {gauss(4,1)};
                \end{axis}
            \end{tikzpicture}
        }
        };
        \node (gauss_hidden3) at (2.8,5.3) [anchor=north west]{
        \resizebox{0.4cm}{!}{
            \begin{tikzpicture}
                \begin{axis}[
                domain=0:8,
                samples=100,
                axis line style={draw=none},
                tick style={draw=none},
                xticklabels=\empty,
                yticklabels=\empty
                ]
                    \addplot[very thick,black,fill=mlred!70] {gauss(4,1)};
                \end{axis}
            \end{tikzpicture}
        }
        };
        \node (gauss_hidden4) at (2.5,3.9) [anchor=north west]{
        \resizebox{0.4cm}{!}{
            \begin{tikzpicture}
                \begin{axis}[
                domain=0:8,
                samples=100,
                axis line style={draw=none},
                tick style={draw=none},
                xticklabels=\empty,
                yticklabels=\empty
                ]
                    \addplot[very thick,black,fill=mlred!70] {gauss(4,1)};
                \end{axis}
            \end{tikzpicture}
        }
        };
        \node (gauss_hidden5) at (3,3.7) [anchor=north west]{
        \resizebox{0.4cm}{!}{
            \begin{tikzpicture}
                \begin{axis}[
                domain=0:8,
                samples=100,
                axis line style={draw=none},
                tick style={draw=none},
                xticklabels=\empty,
                yticklabels=\empty
                ]
                    \addplot[very thick,black,fill=mlred!70] {gauss(4,1)};
                \end{axis}
            \end{tikzpicture}
        }
        };
        \node (gauss_hidden6) at (2.8,4.3) [anchor=north west]{
        \resizebox{0.4cm}{!}{
            \begin{tikzpicture}
                \begin{axis}[
                domain=0:8,
                samples=100,
                axis line style={draw=none},
                tick style={draw=none},
                xticklabels=\empty,
                yticklabels=\empty
                ]
                    \addplot[very thick,black,fill=mlred!70] {gauss(4,1)};
                \end{axis}
            \end{tikzpicture}
        }
        };
        \node (imag2) at (2.4,2.5) [anchor=north west]{
        \begin{tikzpicture}[shorten >=1pt,->,draw=black!50]
            \tikzstyle{every pin edge}=[<-,shorten <=1pt]
            \tikzstyle{neuron}=[circle,minimum size=17pt,inner sep=3pt,draw=black]
            \tikzstyle{weight}=[rectangle,minimum size=17pt,inner sep=0pt,draw=black]
            \node[weight,pin={[pin distance=0.5cm]left:$x$},fill=white] (weights) at (0,0) {W};
            \node[weight,fill=white] (bias) at (0,-1.5) {b};
            \node[neuron,pin={[pin edge={->},pin distance=0.5cm]right:$y$},fill=white] (activation) at (1.5,-0.75) {$\Sigma|\sigma$};
            \path (weights) edge (activation);
            \path (bias) edge (activation);
        \end{tikzpicture}
        };
        \node (gauss_in) at (2.6,2.6) [anchor=north west]{
        \resizebox{0.4cm}{!}{
            \begin{tikzpicture}
                \begin{axis}[
                domain=0:8,
                samples=100,
                axis line style={draw=none},
                tick style={draw=none},
                xticklabels=\empty,
                yticklabels=\empty
                ]
                    \addplot[very thick,black,fill=mlred!70] {gauss(4,1)};
                \end{axis}
            \end{tikzpicture}
        }
        };
        \node (gauss_weight) at (3.2,2.9) [anchor=north west]{
        \resizebox{0.4cm}{!}{
            \begin{tikzpicture}
                \begin{axis}[
                domain=0:8,
                samples=100,
                axis line style={draw=none},
                tick style={draw=none},
                xticklabels=\empty,
                yticklabels=\empty
                ]
                    \addplot[very thick,black,fill=mlred!70] {gauss(4,1)};
                \end{axis}
            \end{tikzpicture}
        }
        };
        \node (gauss_bias) at (3.2,1.4) [anchor=north west]{
        \resizebox{0.4cm}{!}{
            \begin{tikzpicture}
                \begin{axis}[
                domain=0:8,
                samples=100,
                axis line style={draw=none},
                tick style={draw=none},
                xticklabels=\empty,
                yticklabels=\empty
                ]
                    \addplot[very thick,black,fill=mlred!70] {gauss(4,1)};
                \end{axis}
            \end{tikzpicture}
        }
        };
        \node (gauss_out) at (5.5,1.83) [anchor=north west]{
        \resizebox{0.4cm}{!}{
            \begin{tikzpicture}
                \begin{axis}[
                domain=0:8,
                samples=100,
                axis line style={draw=none},
                tick style={draw=none},
                xticklabels=\empty,
                yticklabels=\empty
                ]
                    \addplot[very thick,black,fill=mlred!70] {gauss(4,1)};
                \end{axis}
            \end{tikzpicture}
        }
        };
        \node[align=center] (single_neuron) at (0.1,1.5) [anchor=west]{Single\\ neuron};
        \node (pic_a) at (0,7) [anchor=north west]{a)};
        \node (pic_b) at (0,3) [anchor=north west]{b)};
    \end{tikzpicture}
    };
\end{tikzpicture}

%% file: figures/contour_x.tikz
\begin{tikzpicture}

\begin{groupplot}[%
group style={group size=1 by 2,horizontal sep=4cm,vertical sep=1.5cm},
height=4cm,
width=190pt
]
    \nextgroupplot[%
    colormap/temp,
    colorbar,
    colorbar style={ylabel={$\sigma^\text{GP}_X$},xshift=-0.5cm},
    view={0}{90},
    enlargelimits=false,
    xlabel={$X$},
    ylabel={$S$},
    title=predictive standard deviation
    ]
        \addplot3[surf,shader=interp] table {figures/gp_contour_x_var.txt};
        \addplot3[only marks,mark size=0.8pt] table {figures/gp_train_x.txt};
    \nextgroupplot[%
    colormap/temp,
    colorbar,
    colorbar style={ylabel={$\sigma^\text{BNN}_X$},xshift=-0.5cm},
    view={0}{90},
    enlargelimits=false,
    xlabel={$X$},
    ylabel={$S$}
    ]
        \addplot3[surf,shader=interp] table {figures/pbp_contour_x_var.txt};
        \addplot3[only marks,mark size=0.8pt] table {figures/pbp_train_x.txt};
\end{groupplot}

\end{tikzpicture}

%% file: figures/miscal.tikz
\begin{tikzpicture}

\pgfplotstableread{figures/miscal.txt}\miscalt
\pgfplotstabletranspose{\miscal}\miscalt

\begin{axis}[%
name=first_output,
width=(\linewidth-54.5313pt)/2,
height=3.5cm,
scale only axis,
xmin=0,
xmax=1,
ymin=0,
ymax=1,
domain=0:1,
xlabel style={font=\color{white!15!black}},
xlabel={predicted frequency},
ylabel style={font=\color{white!15!black}},
ylabel={empirical frequency},
title=biomass $X$,
axis background/.style={fill=white},
legend style={at={(0,1)},anchor=north west,legend cell align=left,align=left,draw=white!15!black}
]    
    \addplot[thick,nominal,forget plot] {x};
    \addplot[thick,gp,line legend] table[x expr=\thisrowno{0}/100,y expr=\thisrowno{1}] {\miscal};
    \addplot[thick,bnn,line legend] table[x expr=\thisrowno{0}/100,y expr=\thisrowno{3}] {\miscal};
    \legend{GP,BNN};
\end{axis}

\begin{axis}[%
name=second_output,
at=(first_output.right of south east),
anchor=left of south west,
width=(\linewidth-54.5313pt)/2,
height=3.5cm,
scale only axis,
xmin=0,
xmax=1,
ymin=0,
ymax=1,
domain=0:1,
xlabel style={font=\color{white!15!black}},
xlabel={predicted frequency},
yticklabels={,,},
title=substrate $S$,
axis background/.style={fill=white}
]
    \addplot[thick,nominal] {x};
    \addplot[thick,gp] table[x expr=\thisrowno{0}/100,y expr=\thisrowno{2}] {\miscal};
    \addplot[thick,bnn] table[x expr=\thisrowno{0}/100,y expr=\thisrowno{4}] {\miscal};
\end{axis}

\end{tikzpicture}

%% file: figures/smpc_hybrid.tikz
\begin{tikzpicture}[%
remember picture, node distance=1,scale=1,every node/.style={scale=1}]

\pgfplotstableread{figures/hybrid_smpc_x.txt}\smpchybridt
\pgfplotstabletranspose{\smpchybrid}\smpchybridt

\begin{axis}[%
width=\linewidth-8.83691pt,
height=6cm,
xmin=0,
xmax=50,
xlabel={time in [h]},
ylabel={$X$ in [mg/L]},
axis background/.style={fill=white},
legend style={at={(1,0)},anchor=south east,legend cell align=left,align=left,draw=white!15!black}
]
    \addplot[thick,gp,line legend] table[x expr=\thisrowno{1},y expr=\thisrowno{2}] {\smpchybrid};
    \addplot[thick,bnn,line legend] table[x expr=\thisrowno{1},y expr=\thisrowno{6}] {\smpchybrid};
    \addplot[thick,constraints,line legend] table[x expr=\thisrowno{1},y expr=\thisrowno{4}] {\smpchybrid};
    \addplot[thick,nominal,line legend] table[x expr=\thisrowno{1},y expr=\thisrowno{8}] {\smpchybrid};
    \addplot[thick,nominal] table[x expr=\thisrowno{1},y expr=\thisrowno{9}] {\smpchybrid};
    
    \legend{stochastic hybrid MPC (GP),stochastic hybrid MPC (BNN),hybrid MPC,constraints};

    \coordinate (insetPosition) at (rel axis cs:0.98,0.38);
\end{axis}

\begin{axis}[%
at={(insetPosition)},
anchor={outer south east},
xmin=32,
xmax=42,
ymin=1021,
ymax=1031,
yticklabels={,,},
xticklabels={,,},
tiny
]
    \addplot[thick,gp] table[x expr=\thisrowno{1},y expr=\thisrowno{2}] {\smpchybrid};
    \addplot[thick,bnn] table[x expr=\thisrowno{1},y expr=\thisrowno{6}] {\smpchybrid};
    \addplot[thick,constraints] table[x expr=\thisrowno{1},y expr=\thisrowno{4}] {\smpchybrid};
    \addplot[thick,nominal] table[x expr=\thisrowno{1},y expr=\thisrowno{8}] {\smpchybrid};
\end{axis}
\vspace{-5ex}
\end{tikzpicture}

%% file: figures/comp_time.tikz
\begin{tikzpicture}

\begin{axis}[%
width=\linewidth-8.83691pt,
height=5cm,
enlarge x limits=false,
ymin=0,
ymax=1200,
xlabel=training points,
ylabel={computation time in [s]},
axis background/.style={fill=white},
legend style={at={(0,1)},anchor=north west,legend cell align=left,align=left,draw=white!15!black}
]
    \addplot[gp,only marks,mark size=0.75pt] table {figures/gp_comp_time.txt};
    \addplot[thick,bnn,line legend,domain=0:350] {226.859};

    \legend{stochastic hybrid MPC (GP),stochastic hybrid MPC (BNN)};
\end{axis}

\end{tikzpicture}